\documentclass[twocolumn]{aastex63}

\received{}
\revised{}
\accepted{}
\submitjournal{ApJ Letters}

\shorttitle{RGS Observations of Tycho's SNR}
\shortauthors{Williams et al.}
\graphicspath{{./}{figures/}}

\begin{document}

\title{RGS Observations of Ejecta Knots in Tycho's Supernova Remnant}

\correspondingauthor{Brian J. Williams}
\email{brian.j.williams@nasa.gov}

\author[0000-0003-2063-381X]{Brian J. Williams}
\affiliation{NASA Goddard Space Flight Center, X-ray Astrophysics Laboratory, Greenbelt, MD 20771}

\author{Satoru Katsuda}
\affiliation{Saitama University}

\author{Renata Cumbee}
\affiliation{NASA Goddard Space Flight Center / University of Maryland}

\author{Robert Petre}
\affiliation{NASA Goddard Space Flight Center}

\author{John C. Raymond}
\affiliation{Center for Astrophysics, Harvard \& Smithsonian}

\author{Hiroyuki Uchida}
\affiliation{Kyoto University}



\begin{abstract}

We present results from {\it XMM-Newton/RGS} observations of prominent knots in the southest portion of Tycho's supernova remnant, known to be the remnant of a Type Ia SN in 1572 C.E. By dispersing the photons from these knots out of the remnant with very little emission in front of or behind them, we obtained the nearly uncontaminated spectra of the knots. In the southernmost knot, the RGS successfully resolved numerous emission lines from Si, Ne, O He$\alpha$ and Ly$\alpha$, and Fe L-shell. This is the first clear detection of O lines in Tycho's SNR. Line broadening was measured to be $\sim 3$ eV for the O He$\alpha$ and $\sim 4.5$ eV for Fe L lines. If we attribute the broadening to pure thermal Doppler effects, then we obtain kT$_{O}$ and kT$_{Fe}$ to be $\sim 400$ keV and 1.5 MeV, respectively. These temperatures can be explained by heating in a reverse shock with a shock velocity of $\sim 3500$ km s$^{-1}$. The abundances obtained from fitting the RGS and MOS data together imply substantially elevated amounts of these materials, confirming previous studies that the knots are heated by a reverse shock, and thus contain ejecta material from the supernova. We are unable to find a Type Ia explosion model that reproduces these abundances, but this is likely the result of this knot being too small to extrapolate to the entire remnant.

\end{abstract}



\section{Introduction} 
\label{intro}

Tycho's supernova remnant (SNR), the expanding remains of the supernova of 1572 C.E., is one of the most well-studied SNRs in the sky. First classified as a ``Type I" event by \citet{baade45}, the nature of the event was confirmed as a ``normal" Type Ia SN through light echo studies by \citet{rest08} and \citet{krause08}. In a previous paper \citep{williams17}, we used a combination of X-ray proper motion studies with {\it Chandra} over a long baseline and line-of-sight Doppler velocity measurements from {\it Chandra} spectroscopy to measure the three-dimensional expansion velocities of about 5 dozen knots of emission seen in the remnant (presumed to be ejecta from the SN heated by passage through the reverse shock). We found a mean expansion velocity of these knots of $\sim 4400$ km s$^{-1}$, and found no evidence for any asymmetries in the explosion.

An open question in SNR physics is the amount of heating that ions and electrons experience when hit with a fast shock wave. The degree of ion-electron temperature equilibration must clearly fall somewhere in between two extremes: $T_{i}/T_{e}$ could be as high as $m_{i}/m_{e}$, the ratio of the mass of an ion to the mass of an electron, or it could be as low as unity, if the ions and electrons quickly equilibrate in the postshock gas. These two extremes, however, are several orders of magnitude apart (a factor of 1836 just for protons and electrons, with the factor increasing for heavier ions). Measuring the electron temperature is somewhat straightforward, and is most easily accomplished in young remnants by modeling the shape of the X-ray continuum with any number of models that reproduce emission from an optically-thin plasma. For young SNRs, the electron temperatures are generally of order 1 keV. 

Measuring the ion temperatures is much harder, as this generally requires spectrally-resolving the width of the emission lines that a given ion produces. For the simplest case of a hydrogen ion, it is possible to measure this via the broad component of the H$\alpha$ line at 656 nm. \citet{ghavamian07} compiled broad component widths for several Galactic and Magellanic Cloud SNRs, finding a general trend of increasing $T_{p}/T_{e}$ with increasing shock speed (roughly in proportion to $v^{2}$). Ultraviolet emission from the intermediate ions He II, C IV, N V and O VI has been observed from the narrow ionization zones just behind the shocks in a few SNRs, and they show a similar trend of complete thermal equilibration in slower shocks and $T_{i}$ proportional to m$_{i}$ in faster shocks \citep{raymond2017}. Farther behind the shocks, the metal ions in the X-ray emitting plasma have generally been stripped of electrons to the helium- or hydrogen-like state, meaning their emission lines lie in the X-ray band. However, the current generation of X-ray telescopes such as {\it Chandra} and {\it XMM-Newton} lack the spectral resolving power in their CCD instruments to resolve these lines. Both telescopes are equipped with X-ray grating spectrometers optimized for observing point sources, but analysis of extended sources with gratings is nearly impossible, since the spatial and spectral dimensions become convoluted when the photons are dispersed by the grating. 

Despite this difficulty, grating spectra have been used with some success for SNRs. \citet{vink03} focused on a bright knot in the NW rim of SN 1006, reporting a line width for the 0.57 keV O line of 3.4 $\pm 0.5$ eV, corresponding to an O temperature of 528 $\pm 150$ keV. \citet{broersen13} later used much deeper RGS observations to refine this measurement to a line width of 2.4 $\pm 0.3$ eV. \citet{katsuda13} found narrow O K lines in fast-moving knots in Puppis A, and report an upper limit on the width of 0.9 eV, corresponding to an upper limit on the temperature of 30 keV.  This is much lower than the NW knot in SN 1006, suggesting that these knots are ejecta heated by a slower reverse shock. \citet{miceli19} used {\it Chandra} HETG observations of SN 1987A, in the process of transitioning from a SN to an SNR, to measure the ion temperatures of Ne, Mg, Si, and Fe, finding approximately mass proportional heating for those elements. \citet{kosenko08} and \citet{williams11} used RGS observations of two small, young remnants in the Large Magellanic Cloud to measure the line widths of O and Fe. \citet{uchida19} found an anomalously high ratio of the forbidden-to-resonance O VII lines in a bright knot in the Cygnus Loop, a likely indicator of charge exchange.

\begin{figure}[htb]
\vspace{5mm}
\begin{center}
\includegraphics[width=8cm]{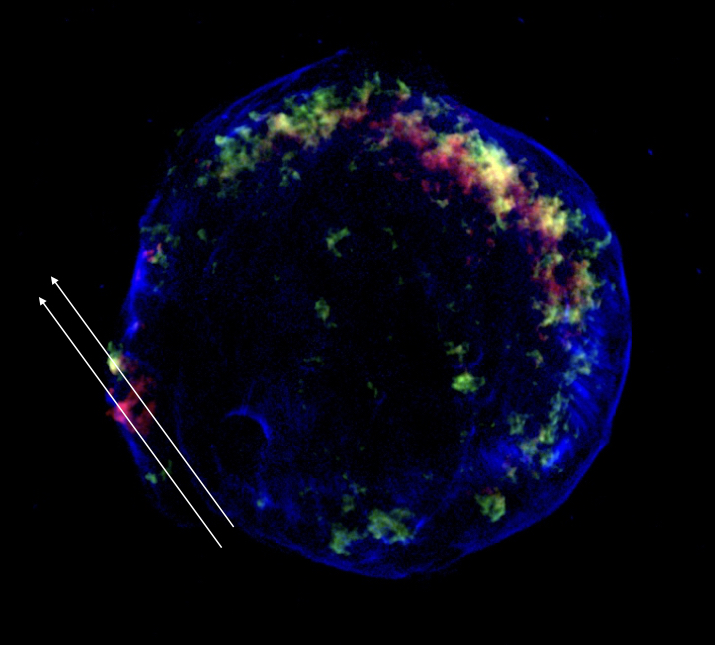}
\end{center}
\caption{A {\it Chandra} image of Tycho with Fe L-shell emission ($0.7-1.0$ keV) in red, Si K$\alpha$ ($1.7-2$ keV) in green, and synchrotron continuum ($4-6$ keV) in blue, superposed with arrows showing the dispersion angle for the RGS observation of the SE knots. The northernmost knot is more Si rich; the Fe-rich knot is southernmost. The intensity scaling is linear, highlighting the high surface brightness of the two knots with respect to their surroundings.}
\label{tycho}
\end{figure}

A particularly bright knot, located along the southeast periphery of Tycho's SNR, has been studied by several authors. \citet{decourchelle01} first reported on this region, finding elevated abundances of intermediate-mass elements. \citet{miceli15} report strong Fe K$\alpha$ emission from this knot, as well as weak emission from other Fe-group elements, such as Cr and Mn. \citet{yamaguchi17} analyzed {\it Suzaku} observations of the knot, instead finding no evidence for emission from those Fe-group elements, and using the upper limits to rule out a Chandrasekhar-mass white dwarf as the progenitor for Tycho's SNR. High spatial resolution images of the knot show it to be a grouping of several knots \citep{yamaguchi17}, with two knots in particular being quite bright. In Figure~\ref{tycho}, we show a three-color {\it Chandra} X-ray image of the remnant, with Fe L-shell and Si K$\alpha$ emission highlighted in red and green, respectively.

In this work, we report on {\it XMM-Newton} Reflection Grating Spectrometer (RGS) observations of these knots. Because they are located on the edge of the remnant, we were able to specify the roll-angle of the telescope in such a way that the light from these knots is dispersed in a nearly tangential direction (see Figure~\ref{tycho}). The emission from the remnant behind these two knots along this dispersion direction is quite weak, meaning that the spectra for these two knots that we extract is very nearly "pristine" (uncontaminated by other emission from the remnant). Being at the edge of the remnant also minimizes the amount of line broadening due to bulk motion from the approaching and receding sides of the remnant. Using the high-resolution RGS spectra, we can resolve the widths of the lines from several atomic species, directly measuring the ion temperatures for O, Ne, Si, and Fe. Further, we can measure the abundances of these elements via joint fits to the RGS and MOS spectra and compare them to various Type Ia nucleosynthesis models. 

\section{Observations}
\label{obs}

We obtained 157 ks of {\it XMM-Newton} observations of Tycho from 2017 Aug 4 to Aug 12. The observations were split into 4 exposures, each of $\sim 40$ ks duration (Obs IDs: 0801840201, 0801840301, 0801840401, and 0801840501). We fixed the roll angle of the telescope to a position angle of 40$^{\circ}$, such that the light from the SE knots would disperse out of the remnant, with a minimal contribution from any emission behind them (see Figure~\ref{tycho}).

We use the first- and second-order RGS data in energy ranges of 0.4--1.5\,keV (31--8.3\,\AA) and 0.7--1.5\,keV (17.7--8.3\,\AA), respectively.  Dividing the energy band into four sections (0.4--0.7\,keV, 0.7--0.9\,keV, 0.9--1.2\,keV, and 1.2--1.5\,keV), we generate RGS responses using the same energy-band images taken with {\it Chandra}, following the analysis of Puppis A \citep{Katsuda12,katsuda13}.  The background is taken from the off-source 1$^{\prime}$-width region of the same observations.  We also analyze MOS data in the energy range of 1.5--10\,keV.  All the raw data were processed using version 16.1.0 of the XMM Science Analysis Software and the calibration data files available in 2017 September.

\section{Results \& Discussion}
\label{results}

\begin{figure}[htb]
\begin{center}
\includegraphics[width=8cm]{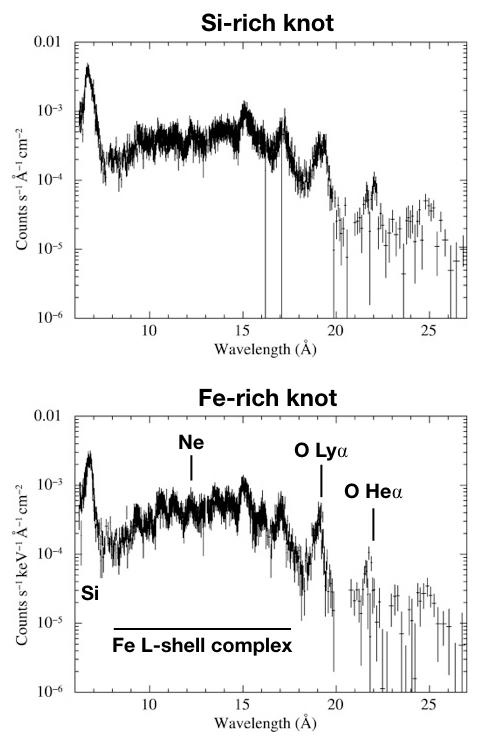}
\end{center}
\caption{The RGS spectra of the two knots identified in the text. For plotting purposes, we only show one RGS order. Lines are identified.}
\label{spectra}
\end{figure}

The RGS spectra of the two knots are shown in Figure~\ref{spectra}. As can be seen solely from Figure~\ref{tycho}, the knots have a somewhat different composition, with the northernmost knot being more Si-rich and the southernmost knot being Fe-rich (``Si-rich'' and ``Fe-rich'' are meant with respect to each other; both knots contain both Si and Fe, prevalent across the entire remnant). The strong Si K$\alpha$ line is clearly present in both knots, as are both the 21.6 \AA\ (0.57 keV) and 18.97 \AA\ (0.65 keV) O lines. {\it This is the first clear detection of oxygen in Tycho's SNR.} While some Fe is present in the Si-rich knot, as is perhaps a weak line from Ne, the better signal-to-noise ratio for all lines is found in the Fe-rich knot. We thus concentrate our analysis for this paper on that knot.

\subsection{Ejecta Abundances}

If the elements seen in the knot correspond to the ejecta from the SN, then the overall abundances can be traced directly back to the progenitor system. To determine the abundances of various heavy elements in the knot, we combine our RGS observations with EPIC-MOS spectra extracted from an identical region. We jointly fit all data (RGS 1\&2 from 0.4-1.5 keV and MOS 1\&2 from 1.5-10 keV) with a single model consisting of several components: a thermal model with a temperature corresponding to the O, Ne, and Mg, another thermal model with a separate temperature corresponding to the intermediate-mass elements, a third component with a separate temperature for the Fe emission, and finally a power-law component to account for the underlying nonthermal synchrotron continuum.

We fit the X-ray spectrum with an absorbed, three {\tt vvnei}s + {\tt power-law} + two Gaussian components model in the XSPEC package \citep{arnaud96}.  The three {\tt vvnei} and {\tt power-law} components are supposed to be thermal and nonthermal synchrotron radiation, respectively.  We multiply a single {\tt gsmooth} model for the thermal components to take account of line-broadening effects, where we take a single line broadening in proportion to the line energy.  We assume that the three {\tt vvnei} components represent SN ejecta, following previous X-ray studies \citep{decourchelle01,yamaguchi17}.  

\begin{figure}[htb]
\begin{center}
\includegraphics[width=9.5cm]{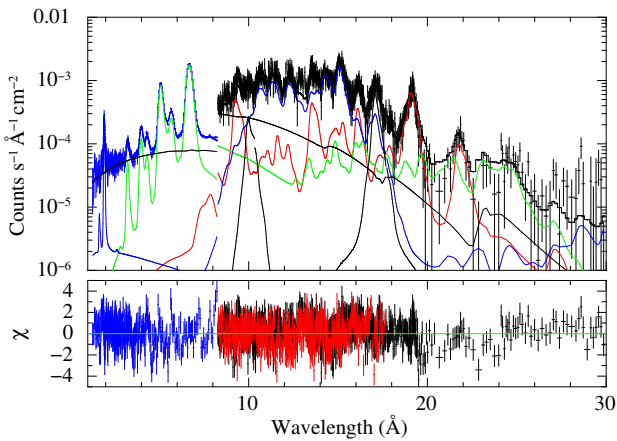}
\end{center}
\caption{Blue and black data points represent MOS 1+2 and RGS 1+2 (1st order), respectively. The three NEI model components, as described in the text, are shown as red (O, Ne, Mg), green (Si, S, Ar, Ca), and blue (Fe) solid lines, while the power-law is a solid black line.}
\label{spectralfit}
\end{figure}

The three {\tt vvnei} components represent ejecta that is (1) O, Ne, and Mg-rich, (2) Si, S, Ar, and Ca-rich, and (3) Fe-rich.  For each component, abundances of these elements are allowed to vary freely, whereas all the other elements including H are fixed to zero, with an exception of the abundance of Ni in the Fe-rich component tied to that of Fe.  To measure relative abundances among the metals, we fix the Si abundance at a certain value (which is insensitive to the relative abundances derived), and tie normalizations of the three {\tt vvnei}s.  The two Gaussians are introduced to compensate for Fe L lines that are not well represented in the {\tt vvnei} model. The temperatures and ionization timescales for each component are allowed to vary independently. The best-fit electron temperatures for the various components were $0.56^{+0.10}_{-0.12}$, $0.99^{+0.02}_{-0.04}$, and $6.44^{+0.29}_{-0.35}$ keV for components 1, 2, and 3, respectively. The ionization timescales ranged from $1-4 \times 10^{10}$ cm$^{-3}$ s. The spectrum and the resulting three-component fit are shown in Figure~\ref{spectralfit}.

\begin{figure}[htb]
\begin{center}
\includegraphics[width=8cm]{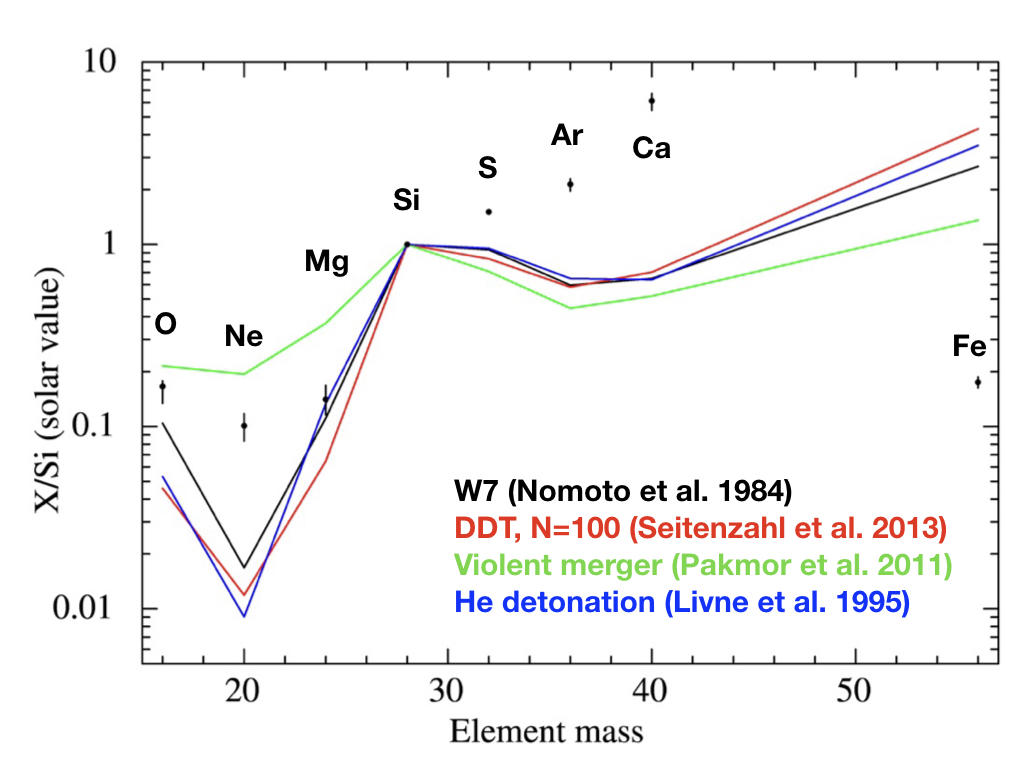}
\end{center}
\caption{Measured abundances of various elements from our joint {\it XMM-Newton} RGS and EPIC modeling, as described in the text, along with theoretical predictions from several Type Ia SN models in the literature.}
\label{abundances}
\end{figure}

We show the measured abundances in Figure~\ref{abundances}, along with several theoretical predictions from various Type Ia SN models. We consider the canonical W7 model from \citet{nomoto84}, the delayed-detonation model (with 100 ignition points) from \citet{seitenzahl13}, a violent merger model from \citet{pakmor11}, and a He-detonation model from \citet{livne95}. None of the models fit particularly well, but there may be several reasons for this. Just because the knots are more consistent with an ejecta origin does not mean there is not {\em any} contribution from forward shocked ISM. Ne, in particular, is not expected in significant quantities in SN Ia ejecta. Additionally, since this is such a small region, extrapolating the abundances we measure here to the remnant as a whole is not a straightforward matter.

\subsection{Line Widths}

Because the signal from the individual lines, particularly the Fe L-shell lines, was not very strong, we first fit a single line width with a linear energy dependence to all lines in the spectrum. We find a best fit line-broadening of 29.5 $\pm 2.1$ eV at 6 keV (though the RGS spectra do not extend to 6 keV, this is where {\it XSpec} pegs the line width in Gaussian-broadening models). The uncertainty represents the 90\% confidence interval. This corresponds to a width of $\sim 5$ eV at 1 keV. When converted to an ion temperature, this leads to the following ion temperatures: kT$_{\rm O}$ = 0.4 $\pm\ 0.05$ MeV, kT$_{\rm Ne}$ = 0.5 $\pm\ 0.07$ MeV, kT$_{\rm Si}$ = 0.7 $\pm\ 0.10$ MeV, and kT$_{\rm Fe}$ = 1.4 $\pm\ 0.19$ MeV. 

If we apply the Rankine-Hugoniot shock jump conditions, such that kT = $\frac{3}{16}m_{i}v_{s}^{2}$ and assume a ``standard" value for the adiabatic index of $\gamma = \frac{5}{3}$, we can conclude that the knots are heated by a reverse shock with a velocity of $\sim 3500$ km s$^{-1}$.

We did attempt to individually fit lines from a given atomic species with the Gaussian-broadening models. For the Fe-L shell lines between 15-17.2 \AA\, we include six Gaussian components (with the widths tied together) to account for the C, D, E, F, G, and H lines from Ne-like Fe (Fe XVII) reported in \citet{gillaspy11}. We obtain a best-fit line width of 3.6 $\pm 0.5$ eV. For the O lines, we include four Gaussian components (again, with the widths tied together) to account for O He-$\alpha$, He-$\beta$, Ly$\alpha$, and Ly$\beta$. We obtain a line width of 4.1 $\pm 0.8$ eV for the oxygen. We were unable to get a reliable individual measurement for Ne as a result of the Ne lines being buried within the Fe L-shell line forest. We obtained a width for Ne of $\sim 20$ eV, which is almost certainly unphysically high. Within errors, the values obtained by fitting individual lines are consistent with those obtained by fitting all lines together. The wide band fit for O result in a best fit value of 3.2 $\pm 0.2$ eV, while the fit for Fe is 3.6 $\pm 0.3$ eV.

There is likely at least some contribution from the bulk velocity of the ejecta material within the knot itself as a result of the reverse shock being driven into it. Quantifying this would require hydrodynamic simulations that are beyond the scope of this work, but averaged over, for instance, a spherical knot, the line-of-sight component is a modest fraction of the reverse shock velocity. When added in quadrature to the thermal width, we expect that the overestimate on the temperature should be modest as well. There may also be contributions from fainter, underlying knots that could contribute to a line broadening. 

The combination of the $T_{e}$ values for the O, Si, and Fe-rich components from the
spectral fits with the ion temperatures assuming a single $\Delta E/E$ value for the line
widths, suggests that $T_{e}/T_{i}$ is about 0.0013 for the lower Z material and 0.005
for Fe. This is generally in keeping with the values of $T_{e}$/$T_{i}$ around 0.003 that
\citet{yamaguchi14} derived from the K$\alpha$ and K$\beta$ lines of iron from the
reverse shock of Tycho for an ambient density near $10^{-24}~\rm g~cm^{-3}$.

However, there is apparently a difference between the shocks in low-Z and high-Z material.
The individual fits for the O and Fe lines are consistent with the equal velocity widths, but
the best fit would increase $T_{O}$ by a factor of 2, reducing $T_{e}$/$T_{O}$ to less than 0.001.
It is likely that the fits determine $T_{e}$ for the Fe component
from the MOS data near 6 keV, and are therefore sensitive to electrons with those energies,
while $T_{e}$ for the O-rich component is fixed by the K$\beta$ to K$\alpha$ ratios of O VII
and O VIII, and is therefore sensitive to electrons below 1 keV.  Either a range of temperatures
in the emitting gas or a non-Maxwellian electron velocity distribution (X-ray synchrotron emission
is present in the region) could produce such a selection effect.

\begin{figure}[htb]
\begin{center}
\includegraphics[width=8cm]{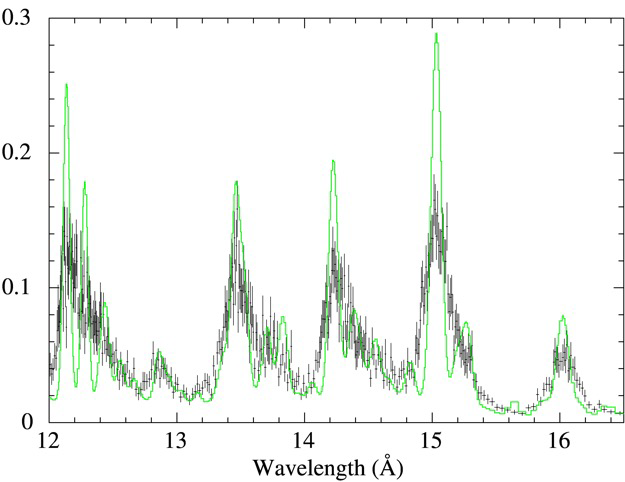}
\end{center}
\caption{A 50 ks XRISM simulation of the SE portion of Tycho's SNR. A ``zero-width" line model is overplotted, as described in the text.}
\label{xrism}
\end{figure}

Further progress in this area will require the use of non-dispersive X-ray spectrometers. The short-lived {\it Hitomi} mission showed the power of such instruments for extended sources, such as the Perseus Cluster \citep{hitomi16}. In Figure~\ref{xrism}, we show a simulation of a 50 ks observation using the {\it X-ray Imaging and Spectroscopy Mission (XRISM)}, scheduled to launch in 2022. This simulation shows only the portion of the spectrum containing the Ne line at 12.1 \AA\ and a series of Fe L-shell lines between 13 and 16~\AA.\ Due to {\it XRISM's} limited spatial resolution, this simulation comes from a larger region ($\sim 1'$ in diameter), which produces much higher signal-to-noise.  

We overplot a model on the simulation that is intentionally a poor fit: the model shows what the lines would look like if they had zero width (this is effectively showing the inherent spectral resolving power of the Resolve instrument on {\it XRISM}). {\it XRISM} will easily resolve the lines we report in this paper, and more importantly, will do so for all lines in the 0.5-12 keV range, including the Fe K$\alpha$ line at 6.4 keV, as well as the S, Ar, and Ca lines shown in Figure~\ref{abundances}, and the weaker Fe-group elements Mn, Cr, and Ni. Because microcalorimeters have a constant energy resolution ($< 7$ eV for {\it XRISM}), the discrepancy between the ``zero-width" model shown in Figure~\ref{xrism} and the measured line widths will be even greater for the lines at higher energies. Looking beyond {\it XRISM}, missions like and {\it Athena} and {\it Lynx} (particularly with their better spatial resolution than {\it XRISM}, important for remnants like Tycho with a complex spatial morphology) will be vital for the study of SNRs, both for measuring the ion temperatures and the abundances across the entire remnant, allowing for a more direct comparison with nucleosynthesis models.

\section{Conclusions}
\label{conclusions}

We report the first direct measurement of heavy-element temperatures in Tycho's SNR. By carefully specifying the roll angle of the telescope, we were able to ensure that the emission from two bright knots on the SE rim of the remnant is nearly uncontaminated by photons from elsewhere in the object. Lines from O are clearly detected for the first time in this remnant. Additionally, we detect lines from Ne, Fe, and Si, all of which are resolved by the RGS instrument. The inferred abundances of these elements are super-solar, consistent with an origin in the SN ejecta. The widths of these lines are well-fit by a Gaussian broadening model with $\sigma = 29.5 \pm 2.1$ eV at 6 keV, which leads to ion temperatures ranging from 400 keV for O to 1.4 MeV for Fe. These temperatures are consistent with reverse-shock heated ejecta.

%

\vspace{5mm}
\facilities{XMM(RGS)}


\software{XSpec \citep{arnaud96}}


\end{document}